%% file: main.tex
\documentclass[remotesensing,article,submit,pdftex,moreauthors]{Definitions/mdpi}
\firstpage{1}
\pubvolume{1}
\issuenum{1}
\articlenumber{0}
\pubyear{2026}
\copyrightyear{2026}
\datereceived{}
\daterevised{}
\dateaccepted{}
\datepublished{}
\doinum{}

\Title{Multi-Fidelity Emulation of Atmospheric Correction Coefficients with Physics-Guided Kolmogorov-Arnold Networks}

\Author{Md Abdullah Al Mazid $^{1}$ and Naphtali Rishe $^{1,*}$}

\AuthorNames{Md Abdullah Al Mazid and Naphtali Rishe}

\address{%
$^{1}$ \quad Knight Foundation School of Computing and Information Sciences, Florida International University, Miami, FL 33199, USA; \texttt{mmazi007@fiu.edu}; \texttt{rishen@fiu.edu}}

\corres{Correspondence: \texttt{rishen@fiu.edu}}
\abstract{Atmospheric correction is a critical preprocessing step in optical remote sensing, but repeated high-fidelity radiative transfer simulations remain computationally expensive for dense look-up-table generation, sensitivity analysis, retrieval support, and operational preprocessing. This study presents a physics-aware multi-fidelity surrogate framework for emulating atmospheric correction coefficients using paired 6S and libRadtran simulations. Atmospheric and geometric states are sampled using Latin Hypercube Sampling, and both radiative transfer models are evaluated under matched conditions for Sentinel-2 bands using spectral-response-function-aware coefficient generation. The high-fidelity targets are path reflectance, total transmittance, and spherical albedo. A physics-guided Kolmogorov-Arnold Network, termed pKANrtm, receives the atmospheric state and low-fidelity 6S coefficients, predicts the residual relative to libRadtran, and reconstructs the high-fidelity coefficients. The pKANrtm model uses an Efficient-KAN architecture and is trained with a physics-consistency penalty applied in the original coefficient space. The proposed model is evaluated against state-of-the-art regression-based RTM surrogates. Across both standard and out-of-distribution evaluation settings, pKANrtm achieves the strongest overall predictive performance among the compared models. Band-wise analysis shows that most Sentinel-2 bands are accurately emulated, while absorption-sensitive bands remain comparatively challenging. Runtime benchmarking demonstrates substantial acceleration relative to libRadtran, with GPU inference providing approximately four orders of magnitude single-sample speedup and batched inference reaching tens of thousands of samples per second. These results indicate that physics-aware multi-fidelity pKANrtm emulation provides an accurate, physically structured, and computationally efficient strategy for atmospheric correction coefficient generation.}

\keyword{atmospheric correction; radiative transfer model; libRadtran; 6S; multi-fidelity learning; physics-guided learning; Kolmogorov-Arnold networks; surrogate modeling; spectral response function; Sentinel-2; Latin Hypercube Sampling; remote sensing}

\begin{document}

\section{Introduction}
\input{sections/01_introduction}

\section{Related Work}
\input{sections/02_related_work}

\section{Materials and Methods}
\input{sections/03_materials_and_methods}

\section{Experimental Setup}
\input{sections/04_experimental_setup}

\section{Results and Discussion}
\input{sections/05_results_and_discussion}

\section{Conclusions}
\input{sections/06_conclusions}

\authorcontributions{Conceptualization, M.A.A.M.; methodology, M.A.A.M.; validation, N.R.; investigation, M.A.A.M.; data curation, M.A.A.M.; writing-original draft, M.A.A.M.; writing-review and editing, M.A.A.M. and N.R.; visualization, M.A.A.M.; supervision, N.R. All authors have read and agreed to the published version of the manuscript.}

\funding{This material is based in part upon work supported by the National Science Foundation under Grant Nos. CNS-2018611 and CNS-1920182.}

\institutionalreview{Not applicable.}

\informedconsent{Not applicable.}

\dataavailability{The dataset generated and used in this study is publicly available through Hugging Face at \url{https://huggingface.co/datasets/mazid-rafee/pKANrtm}. The source code, model implementation, training scripts, benchmarking utilities, and analysis workflow are publicly available at \url{https://github.com/mazid-rafee/pKANrtm-atmospheric-correction-dl-surrogate}.}

\acknowledgments{The authors acknowledge the support of the Knight Foundation School of Computing and Information Sciences, Florida International University.}

\conflictsofinterest{The authors declare no conflicts of interest.}

\abbreviations{Abbreviations}{
\noindent
\begin{tabular}{@{}ll}
6S & Second Simulation of the Satellite Signal in the Solar Spectrum \\
AOD & Aerosol Optical Depth \\
AOD550 & Aerosol Optical Depth at 550 nm \\
CWV & Column Water Vapour \\
HF & High Fidelity \\
KAN & Kolmogorov-Arnold Network \\
LF & Low Fidelity \\
LHS & Latin Hypercube Sampling \\
MAE & Mean Absolute Error \\
OOD & Out-of-Distribution \\
RMSE & Root Mean Squared Error \\
RTM & Radiative Transfer Model \\
SMAPE & Symmetric Mean Absolute Percentage Error \\
SRF & Spectral Response Function \\
TOA & Top of Atmosphere
\end{tabular}}

\reftitle{References}
\bibliography{references}

\PublishersNote{}
\end{document}

%% file: sections/01_introduction.tex
Atmospheric correction is a fundamental prerequisite for quantitative optical remote sensing because top-of-atmosphere (TOA) radiance or reflectance contains coupled contributions from surface reflection, molecular absorption, aerosol scattering, atmospheric path radiance, illumination geometry, and sensor-view geometry. Physics-based radiative transfer models (RTMs) provide the standard mechanism for representing these interactions and for converting sensor-level observations into surface-reflectance products. Classical and operational atmospheric-correction workflows therefore rely on RTMs and numerical radiative-transfer solvers such as 6S, MODTRAN, libRadtran, DISORT, and SBDART, as well as mission-oriented processors such as LaSRC, ATCOR, LEDAPS-like Landsat workflows, and multi-temporal Sentinel/Landsat correction approaches \citep{vermote1997second,kotchenova2006validation,berk2014modtran6,mayer2005libradtran,stamnes1988disort,ricchiazzi1998sbdart,masek2006landsat,vermote2016lasrc,richter2002atcor,hagolle2015multitemporal,claverie2018hls}.

Despite their physical interpretability, RTM-based workflows remain computationally demanding when simulations must be repeated over large atmospheric, geometric, spectral, or surface-state spaces. This cost is especially important for dense look-up table (LUT) generation, imaging spectroscopy retrieval, uncertainty quantification, global sensitivity analysis, and large-scale operational preprocessing. Benchmarking exercises such as ACIX have shown that atmospheric correction remains a challenging and processor-dependent problem, with differences arising from aerosol and water-vapour retrieval, solver choices, surface assumptions, and spectral configuration \citep{gao2009atmospheric,doxani2018acix,doxani2023acixii,cremer2025acix}. In parallel, optimal-estimation approaches for imaging spectroscopy have demonstrated the value of physically structured observation models, but they also highlight the burden of repeated high-fidelity forward simulations \citep{thompson2018optimal,brodrick2021generalized}.

Surrogate modeling has emerged as a practical response to this bottleneck. Instead of executing a full RTM for every query, an emulator learns the relationship between atmospheric-geometric inputs and selected RTM outputs. Early remote-sensing emulators used Gaussian processes to approximate RTMs while also providing uncertainty and derivative information \citep{gomezdans2016efficient,verrelst2016emulation}. Subsequent work extended RTM emulation to MODTRAN atmospheric-correction LUTs, neural-network radiative-transfer approximation for imaging spectroscopy, dense LUT generation through sRTMnet, and surrogate modeling for trace-gas retrievals \citep{servera2021systematic,bue2019neural,brodrick2021generalized,brence2023surrogate}. Recent studies and reviews further show that emulators can accelerate RTM workflows by orders of magnitude, including in weather and climate radiation parameterization, although accuracy, extrapolation, uncertainty, sampling design, and physical consistency remain central concerns \citep{pal2019surrogate,liu2020radnet,roh2020nnradiation,aghdaminia2024modtran,basener2023gpdlac,verrelst2025rtm}.

A related opportunity is multi-fidelity emulation. Multi-fidelity modeling combines inexpensive low-fidelity simulations with fewer or more expensive high-fidelity simulations, allowing the learned model to exploit correlations between simulators rather than learning the high-fidelity response from raw inputs alone \citep{kennedy2000multifidelity,forrester2007multifidelity,perdikaris2017nonlinear,peherstorfer2018survey}. This idea is especially attractive for atmospheric correction because a faster RTM can often capture much of the broad physical structure, while a more expensive RTM provides a more detailed target under the chosen configuration. The sRTMnet framework illustrates this principle by correcting a reduced-order/surrogate radiative-transfer calculation toward high-fidelity MODTRAN outputs \citep{brodrick2021generalized}. In this paper, we adopt a related philosophy using 6S as a low-fidelity RTM and libRadtran as the high-fidelity target.

The target representation is another important design choice. Many previous studies emulate spectra, radiances, transmittance profiles, or surface reflectance directly \citep{bue2019neural,stegmann2022fastrt,lagerquist2021dlrtm,yao2023physicsdlrt}. Here, we instead focus on atmospheric-correction coefficients: path reflectance, total transmittance, and spherical albedo. These quantities are compact, physically meaningful, and directly connected to standard Lambertian atmospheric-correction equations. Coefficient-level emulation therefore provides a middle ground between full RTM execution and purely image-to-image surface-reflectance prediction: the learned model accelerates the physics-based correction pipeline while preserving its interpretable coefficient structure.

This study further explores Kolmogorov-Arnold Network (KAN) family models for coefficient-level RTM emulation. KANs replace fixed nodal activation functions with learnable univariate functions on network edges, offering an alternative nonlinear function-approximation mechanism to conventional multilayer perceptrons \citep{liu2024kan}. Because atmospheric correction coefficients are governed by smooth but strongly nonlinear interactions among aerosol loading, water vapour, ozone, geometry, elevation, and wavelength, KAN-family models are a natural candidate for compact scientific regression. We also evaluate a physics-guided KAN variant, pKANrtm, motivated by broader work showing that physical constraints and consistency penalties can improve the credibility of learned scientific surrogates \citep{raissi2019pinn,karniadakis2021physics,yao2023physicsdlrt}.

The main contributions of this work are as follows. First, we construct a paired SRF-aware multi-fidelity dataset generation pipeline using Latin Hypercube Sampling to define matched atmospheric and geometric states for Sentinel-2 spectral bands, with 6S as the low-fidelity RTM and libRadtran with the DISORT solver as the high-fidelity RTM. Second, we formulate high-fidelity atmospheric-correction coefficient emulation as a residual multi-fidelity learning problem, where 6S coefficients provide structured low-fidelity guidance for predicting libRadtran coefficients. Third, we introduce pKANrtm, a physics-guided Kolmogorov-Arnold Network surrogate that combines Efficient-KAN representation with a physics-consistency penalty applied in the original coefficient space. Fourth, we compare pKANrtm with state-of-the-art regression-based RTM surrogates under both standard and out-of-distribution split regimes. Fifth, we provide a comprehensive evaluation using aggregate accuracy, coefficient-wise error, band-wise error, conditional-error diagnostics, qualitative spectral-case analysis, and runtime benchmarking against RTM baselines. Sixth, we release the dataset and implementation code publicly to support reproducibility, model comparison, and future benchmarking of atmospheric-correction surrogate models. Together, these contributions provide a practical pathway toward fast, accurate, and scalable atmospheric-correction coefficient emulation while retaining the physical structure of an RTM-based workflow.

%% file: sections/02_related_work.tex
\subsection{Physics-based atmospheric correction and radiative-transfer modeling}

Atmospheric correction has traditionally been formulated as an inverse radiative-transfer problem in which atmospheric path effects are removed from TOA observations to retrieve surface reflectance. The physical basis of these methods is the radiative-transfer equation and its numerical solution under assumptions about atmospheric composition, aerosol properties, geometry, and surface reflectance. Widely used RTMs include 6S and its vector-polarized validation lineage, which have been extensively adopted for multispectral atmospheric correction; MODTRAN, which supports detailed atmospheric transmission and radiance simulation; libRadtran, which provides a flexible framework with several radiative-transfer solvers; and SBDART, a research-oriented plane-parallel radiative-transfer code \citep{vermote1997second,kotchenova2006validation,berk2014modtran6,mayer2005libradtran,ricchiazzi1998sbdart}. Solver development, such as DISORT and related discrete-ordinate methods, remains central to accurate multiple-scattering calculations in layered atmospheres \citep{stamnes1988disort}.

Operational atmospheric-correction processors build on these physical models but differ in their assumptions, ancillary inputs, aerosol and water-vapour retrieval strategies, spectral treatment, and correction of adjacency or terrain effects. LaSRC uses 6S-based radiative-transfer calculations for Landsat surface reflectance production, LEDAPS established earlier large-area Landsat surface-reflectance processing, and HLS demonstrates the need for cross-sensor harmonization between Landsat and Sentinel-2 surface reflectance; ATCOR-type workflows combine atmospheric and topographic correction for airborne and satellite imagery \citep{masek2006landsat,vermote2016lasrc,claverie2018hls,richter2002atcor}. For imaging spectroscopy, atmospheric-correction reviews have emphasized the importance of spectral absorption structure, smoothing, and radiative-transfer assumptions, while optimal-estimation methods such as ISOFIT explicitly fit atmospheric and surface states by repeated forward-model evaluation, providing a physically rigorous but computationally demanding framework \citep{gao2009atmospheric,thompson2018optimal}. Community inter-comparisons such as ACIX and ACIX-II have demonstrated that atmospheric correction remains a nontrivial source of uncertainty across processors, sensors, and validation sites \citep{doxani2018acix,doxani2023acixii}. The recent ACIX-III hyperspectral benchmark further emphasizes the need for transparent validation of aerosol, water-vapour, and surface-reflectance outputs in modern imaging spectroscopy workflows \citep{cremer2025acix}. Multi-temporal aerosol retrieval strategies are also important in operational optical correction pipelines because repeated observations can help separate slowly varying surface reflectance from atmospheric variability \citep{hagolle2015multitemporal}.

\subsection{RTM surrogate modeling and emulator design}

RTM surrogate modeling replaces repeated calls to an expensive physical model with a learned approximation trained on RTM-generated samples. Early work by \citet{gomezdans2016efficient} showed that Gaussian-process emulators can approximate radiative-transfer codes and provide uncertainty and derivative information useful for inversion. \citet{verrelst2016emulation} further demonstrated the use of emulators for fast global sensitivity analysis of leaf, canopy, and atmospheric RTMs. These studies helped establish emulator training as a viable way to explore high-dimensional RTM input spaces without exhaustive physical-model execution.

In atmospheric correction, RTM emulators have been used to replace or densify LUT operations. \citet{servera2021systematic} systematically assessed MODTRAN emulators for atmospheric correction, considering the accuracy and computational trade-offs of different emulator designs. \citet{bue2019neural} developed neural-network radiative-transfer approximations for imaging spectroscopy, while \citet{brodrick2021generalized} proposed sRTMnet, a hybrid reduced-order RTM and neural emulator that enables dense LUT generation with high-fidelity MODTRAN-like accuracy. For atmospheric trace-gas retrieval, \citet{brence2023surrogate} showed that dimensionality reduction and surrogate models can substantially accelerate forward simulations. More recent studies have applied fully connected networks, autoencoders, random forests, Gaussian processes, and deep-learning regressors to MODTRAN-like atmospheric correction and compensation problems \citep{basener2023gpdlac,aghdaminia2024modtran}. In atmospheric and climate modeling, neural-network radiation emulators have also shown substantial acceleration for shortwave and longwave radiative-transfer parameterizations, while highlighting generalization and coupled-model stability issues \citep{pal2019surrogate,liu2020radnet,roh2020nnradiation}.

Sampling strategy is a central part of emulator design because the learned model can only generalize over the state space represented during training. Space-filling designs such as Latin Hypercube Sampling are commonly used to cover high-dimensional RTM inputs more efficiently than naive random sampling \citep{mckay1979lhs,gomezdans2016efficient,brence2023surrogate,verrelst2025rtm}. The recent review by \citet{verrelst2025rtm} highlights several persistent issues in RTM emulation: balancing accuracy and computational speed, representing spectral dimensionality efficiently, estimating uncertainty, avoiding poor extrapolation, and maintaining physical interpretability. These issues motivate the present focus on compact atmospheric-correction coefficients, state-level split control, out-of-distribution testing, and physics-guided regularization.

\subsection{Learning-based atmospheric correction and physics-guided surrogates}

Learning-based atmospheric correction has been studied both as a direct image-to-image reflectance prediction problem and as an RTM-assisted emulation problem. \citet{shah2022deepac} organized deep-learning atmospheric correction approaches into physics-supported, physics-aware, and physics-agnostic categories. \citet{basener2023gpdlac} compared Gaussian-process and deep-learning atmospheric correction models and emphasized the value of interpretable probabilistic modeling. More recently, \citet{shah2025spatiotemporal} proposed a spatio-temporal atmospheric correction network, showing that temporal information can improve surface-reflectance prediction compared with purely spatial learning. These studies demonstrate the promise of data-driven atmospheric correction, but many of them predict reflectance directly rather than emulating the intermediate coefficients used by physical correction equations.

Physics-guided machine learning offers a complementary strategy by embedding known constraints into the model architecture or training objective. Physics-informed neural networks introduced a general framework for enforcing differential-equation constraints during learning \citep{raissi2019pinn}, while broader reviews of physics-informed machine learning emphasize the importance of consistency, inductive bias, and scientific interpretability \citep{karniadakis2021physics}. In atmospheric radiative transfer, physics-incorporated deep-learning frameworks have encoded conservation relationships among fluxes and heating rates, and other deep-learning RTM accelerators have focused on preserving physically meaningful radiation quantities while reducing computational cost \citep{lagerquist2021dlrtm,stegmann2022fastrt,yao2023physicsdlrt}. The pKANrtm model follows this broader philosophy by preserving supervised residual learning while penalizing physically inconsistent coefficient predictions in the original coefficient space.

\subsection{Multi-fidelity learning and KAN-family models}

Multi-fidelity modeling combines simulations or observations obtained at different levels of cost and accuracy. The classical Kennedy-O'Hagan framework formalized the use of inexpensive approximations to improve prediction of expensive computer-code outputs \citep{kennedy2000multifidelity}, and co-kriging-based multi-fidelity surrogate optimization further demonstrated how multiple analysis fidelities can be combined for efficient model exploration \citep{forrester2007multifidelity}. Later nonlinear multi-fidelity methods showed that information can be fused across fidelities even when the relationship between low- and high-fidelity models is not purely linear \citep{perdikaris2017nonlinear}. Broader surveys of multi-fidelity uncertainty propagation, inference, and optimization emphasize that low-fidelity models are most useful when they are cheap, correlated with the high-fidelity model, and used with validation safeguards \citep{peherstorfer2018survey}. This perspective is particularly relevant to atmospheric correction because fast RTMs such as 6S can provide physically meaningful low-fidelity coefficients, while more expensive models such as libRadtran or MODTRAN can serve as high-fidelity targets under selected configurations. sRTMnet is a closely related example in which a faster radiative-transfer approximation is corrected toward high-fidelity MODTRAN behavior through a neural emulator \citep{brodrick2021generalized}.

From an architectural perspective, multilayer perceptrons remain common baselines for scientific regression and RTM emulation \citep{rumelhart1986learning,taud2018mlp}. KANs provide a newer alternative inspired by the Kolmogorov-Arnold representation theorem, using learnable univariate functions on edges rather than fixed nodal activations \citep{liu2024kan}. Practical implementations such as \texttt{efficient-kan} make KAN-family models easier to evaluate in applied scientific regression settings \citep{blealtan2024efficientkan}. However, KANs have not yet been widely studied for atmospheric RTM emulation. To the best of our knowledge, SRF-aware multi-fidelity emulation of atmospheric-correction coefficients from paired 6S and libRadtran simulations using physics-guided KAN models remains unexplored. This gap motivates the proposed pKANrtm framework.

%% file: sections/03_materials_and_methods.tex
\subsection{Problem formulation}

Let $\mathbf{x}$ denote a state vector containing the atmospheric, geometric, categorical, and spectral descriptors required to run the RTMs. In this study, $\mathbf{x}$ includes solar zenith angle, view zenith angle, relative azimuth angle, aerosol optical depth at 550 nm (AOD550), column water vapour (CWV), ozone amount, surface elevation, atmospheric profile, aerosol setting, and Sentinel-2 band metadata. For each state and band, the low-fidelity RTM produces a coefficient vector $\mathbf{y}_{LF}$ and the high-fidelity RTM produces a coefficient vector $\mathbf{y}_{HF}$:
\begin{equation}
\mathbf{y}=\left[\rho_{\mathrm{path}},\;T_{\mathrm{total}},\;s\right],
\end{equation}
where $\rho_{\mathrm{path}}$ is path reflectance, $T_{\mathrm{total}}$ is total transmittance, and $s$ is spherical albedo.

The surface-reflectance correction equation can be written in coefficient form as
\begin{equation}
\rho_s = \frac{\rho_{TOA}-\rho_{\mathrm{path}}}{T_{\mathrm{total}} + s(\rho_{TOA}-\rho_{\mathrm{path}})},
\end{equation}
where $\rho_{TOA}$ denotes TOA reflectance and $\rho_s$ denotes the retrieved surface reflectance under the Lambertian surface assumption used in this coefficient formulation. Therefore, small coefficient errors can propagate directly into surface-reflectance retrieval, motivating coefficient-wise and band-wise evaluation rather than only aggregate model scores.

The primary learning problem is residual multi-fidelity emulation. Given $\mathbf{x}$ and the low-fidelity coefficient vector $\mathbf{y}_{LF}$, the model learns the high-fidelity residual
\begin{equation}
\mathbf{r}=\mathbf{y}_{HF}-\mathbf{y}_{LF}.
\end{equation}
A trained model $f_{\theta}$ predicts $\widehat{\mathbf{r}}=f_{\theta}(\mathbf{x},\mathbf{y}_{LF})$, and the high-fidelity coefficient estimate is reconstructed as
\begin{equation}
\widehat{\mathbf{y}}_{HF}=\mathbf{y}_{LF}+\widehat{\mathbf{r}}.
\end{equation}
This setting targets the correction from low- to high-fidelity while conditioning on both state and low-fidelity coefficients.

\subsection{Radiative transfer models}

The low-fidelity model is 6S, and the high-fidelity model is libRadtran \citep{vermote1997second,mayer2005libradtran}. 6S is computationally efficient and widely used for atmospheric correction and multispectral remote-sensing simulation. libRadtran provides a flexible radiative-transfer framework with multiple solvers and spectral configurations, and it is used here as the higher-fidelity target model. Both RTMs are evaluated under a shared state manifest so that every accepted row has matched low- and high-fidelity coefficients.

This pairing is not intended to claim that one RTM is universally correct and the other is universally incorrect. Rather, it reflects a practical hierarchy for the present experimental configuration: 6S is used as a fast physical approximation, while libRadtran is treated as the higher-fidelity target whose repeated execution is expensive. The learned model therefore approximates the libRadtran coefficient response conditional on both the physical state and the corresponding 6S output.

\subsection{Latin Hypercube state sampling}

Continuous state variables are sampled using Latin Hypercube Sampling (LHS) to improve coverage of the multidimensional atmospheric and geometric domain \citep{mckay1979lhs,brence2023surrogate,verrelst2025rtm}. The sampled variables include AOD550, CWV, ozone amount, solar zenith angle, view zenith angle, relative azimuth angle, and elevation. Categorical variables such as aerosol model and atmospheric profile are assigned from a shared schema so that both RTMs receive conceptually matched conditions.

Each unique atmospheric-geometric state is assigned a deterministic state identifier. This identifier is used for reproducibility, traceability, and split assignment. Importantly, data splitting is performed at the state level rather than at the row level; all band rows belonging to the same state are assigned to the same train, validation, or test split. This prevents leakage where the same physical state could appear in training for one band and testing for another.

\subsection{SRF-aware Sentinel-2 coefficient generation}

The study uses SRF-aware integration for 13 Sentinel-2 bands: B1, B2, B3, B4, B5, B6, B7, B8, B8A, B9, B10, B11, and B12. For each band, the RTM spectral response is convolved with the corresponding Sentinel-2 spectral response function. The band-level coefficient $c_b$ for a generic spectral coefficient $c(\lambda)$ is computed as
\begin{equation}
c_b=\frac{\int c(\lambda)R_b(\lambda)\,d\lambda}{\int R_b(\lambda)\,d\lambda},
\end{equation}
where $R_b(\lambda)$ is the SRF of band $b$. The same SRF-aware integration is applied to both 6S and libRadtran outputs, ensuring that the multi-fidelity targets are defined on the same sensor-level spectral support.

SRF integration is particularly important near absorption-sensitive regions. A central-wavelength approximation can miss within-band variation in transmittance or path reflectance, whereas SRF-aware integration better represents the quantity observed by the multispectral sensor. This makes the resulting coefficient dataset more relevant for practical Sentinel-2 atmospheric correction workflows \citep{drusch2012sentinel,brodrick2021generalized}.

\subsection{Coefficient recovery and quality control}

For each accepted state-band pair, atmospheric correction coefficients are recovered from RTM simulations at multiple Lambertian surface-reflectance anchors. In the default data-generation configuration, the anchor inputs are $\rho_0=0.0$, $\rho_1=0.2$, and $\rho_2=0.6$. The corresponding RTM-computed TOA reflectance responses are denoted $r_0$, $r_1$, and $r_2$, respectively. Therefore, $r_0$, $r_1$, and $r_2$ are not fixed global constants; they vary with atmospheric state, viewing and illumination geometry, spectral band, and RTM fidelity level. For a given anchor reflectance $\rho_i$, the coefficient model can be written as
\begin{equation}
r_i = \rho_{\mathrm{path}} +
\frac{T_{\mathrm{total}}\rho_i}{1-s\rho_i},
\end{equation}
where $r_i$ is the RTM-computed TOA reflectance at anchor $\rho_i$. The three anchor-response pairs provide sufficient information to estimate $\rho_{\mathrm{path}}$, $T_{\mathrm{total}}$, and $s$ consistently for both fidelity levels. The same recovery procedure is used for 6S and libRadtran so that residual learning compares like with like.

Quality control is applied before model training. Rows are excluded if they contain non-finite values, violate monotonicity expectations over the reflectance anchors, or produce physically implausible coefficient values. This step is important because coefficient-recovery instability is not the same as neural prediction error. The quality-control statistics are retained for interpretation, especially for bands with strong absorption or low usable signal.

After quality control, the final training dataset was constructed as the paired intersection of the 6S and libRadtran outputs. The intersection was performed using the tuple \texttt{(state\_id, band)} so that a sample was retained only when both RTMs produced valid coefficients for the same atmospheric-geometric state and Sentinel-2 band. The resulting dataset, denoted \texttt{qavalid\_intersection\_libradtran\_6s\_50k\_13b}, contains 609,722 paired state-band samples from 50,000 unique atmospheric states. For most Sentinel-2 bands, 50,000 paired samples were retained. The exception is B10, for which 9,722 valid paired samples remained after filtering. This reduction is consistent with the greater coefficient-recovery instability of the cirrus/absorption-sensitive spectral region.

A row was retained only when it satisfied the following quality-control criteria in both RTMs: \texttt{qa\_valid = true}, \texttt{monotonicity\_failed = false}, \texttt{T\_total\_was\_clamped = false}, \texttt{rho\_path\_was\_clamped = false}, and \texttt{spher\_alb\_was\_clamped = false}. These constraints ensure that the learning problem is defined on physically admissible coefficient triplets rather than on artifacts introduced by unstable coefficient recovery.

\begin{table}[H]
\caption{Summary of the QA-valid paired 6S-libRadtran dataset.}
\label{tab:dataset_summary}
\centering
\begin{tabular}{ll}
\toprule
Property & Value \\
\midrule
Dataset pair & \texttt{qavalid\_intersection\_libradtran\_6s\_50k\_13b} \\
Low-fidelity RTM & 6S \\
High-fidelity RTM & libRadtran \\
Sampling strategy & Latin Hypercube Sampling \\
Unique atmospheric-geometric states & 50,000 \\
Intersection key & \texttt{state\_id, band} \\
Final paired state-band samples & 609,722 \\
Bands with 50,000 samples & B1-B9, B8A, B11, B12 \\
B10 valid samples & 9,722 \\
Retained QA criteria & valid, monotonic, unclamped coefficients \\
\bottomrule
\end{tabular}
\end{table}

\subsection{pKANrtm architecture}

\begin{figure}[H]
\centering
\includegraphics[width=\textwidth]{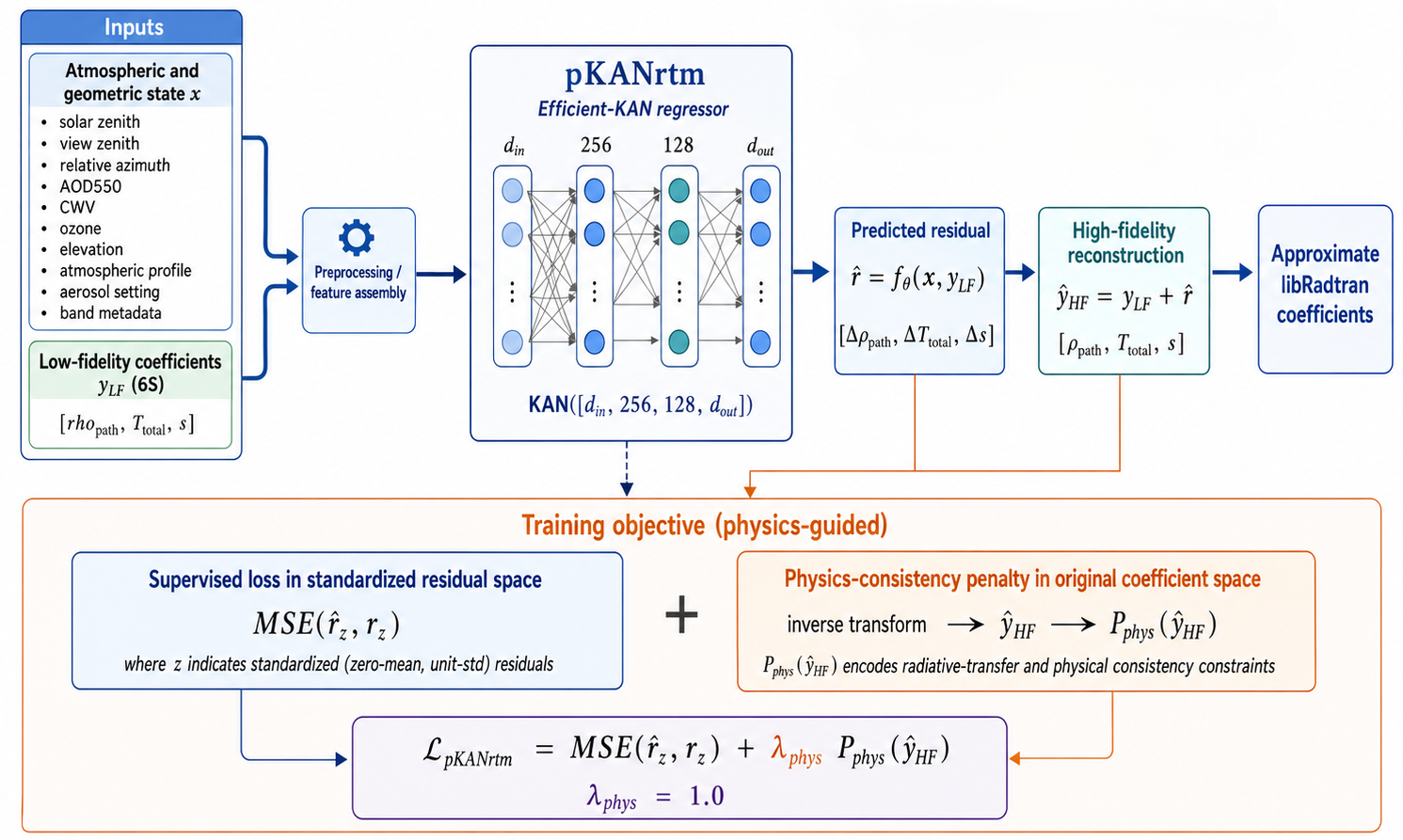}
\caption{Architecture and physics-guided training workflow of pKANrtm}
\label{fig:pkanrtm_architecture}
\end{figure}

The proposed model is a physics-guided Kolmogorov-Arnold Network, referred to as pKANrtm with neural architecture:
\begin{equation}
\mathrm{KAN}\left([d_{in}, 256, 128, d_{out}]\right),
\end{equation}
where $d_{in}$ denotes the number of preprocessed input features and $d_{out}$ denotes the number of target residual coefficients. Thus, pKANrtm is implemented as a two-hidden-layer Efficient-KAN regressor with hidden widths of 256 and 128. The model is implemented using the \texttt{efficient-kan} PyTorch implementation \citep{blealtan2024efficientkan}.

pKANrtm is trained only with supervised mean-squared error (MSE) and an additional physics-guided regularization term. This design separates the effect of the KAN architecture from the effect of physics-guided training. Figure~\ref{fig:pkanrtm_architecture} summarizes the input assembly, residual prediction, high-fidelity coefficient reconstruction, and physics-guided training objective.

\subsection{Physics-guided training objective}

Targets are standardized using a \texttt{StandardScaler} fitted on the training set. The supervised term is computed in standardized target space using MSE. For the physics penalty, predictions are inverse transformed back to the original coefficient scale, and the penalty is computed on the unscaled predicted coefficients. The pKANrtm objective is
\begin{equation}
\mathcal{L}_{pKANrtm}=\mathrm{MSE}(\widehat{\mathbf{r}}_{z},\mathbf{r}_{z})+\lambda_{phys}\,\mathcal{P}_{phys}(\widehat{\mathbf{y}}_{HF}),
\end{equation}
where $\widehat{\mathbf{r}}_{z}$ and $\mathbf{r}_{z}$ denote standardized predicted and true residuals, $\widehat{\mathbf{y}}_{HF}$ is the inverse-transformed reconstructed high-fidelity coefficient vector, and $\mathcal{P}_{phys}$ is the physics-consistency penalty. The default penalty weight is $\lambda_{phys}=1.0$.

The purpose of the physics term is not to replace supervised learning, but to discourage physically inconsistent coefficient predictions after reconstruction in the original coefficient space. This is important because a prediction can appear small in standardized error while still violating coefficient constraints that matter for atmospheric correction.

\subsection{Training protocol}

Models are trained with the Adam optimizer, an initial learning rate of $10^{-3}$, batch size 2048, maximum training length of 100 epochs, and early stopping patience of 20 epochs. The best checkpoint is selected using validation loss. Mixed precision can be enabled when supported, but the reported configuration is defined by the same optimization protocol across model comparisons.

\subsection{Implementation and reproducibility}

The dataset and source code used for this study are made publicly available to support reproducibility and future comparison. The pKANrtm dataset is hosted on Hugging Face, and the model implementation, training scripts, benchmarking utilities, and analysis workflow are provided in the accompanying GitHub repository. The public release enables independent inspection of the data format, model configuration, and experimental pipeline.

%% file: sections/04_experimental_setup.tex
\subsection{Dataset construction and coverage}

The paired dataset is generated by evaluating 6S and libRadtran over the same LHS-sampled state manifest. For each atmospheric and geometric state, each RTM is evaluated over the spectral support required for SRF-aware Sentinel-2 band integration. The libRadtran simulations are performed using the DISORT solver, and the solver metadata, including the number of streams, is retained in the generated records for reproducibility. The stored rows include the state identifier, split label, continuous and categorical input variables, band metadata, 6S coefficients, libRadtran coefficients, and residual targets. This design supports both direct comparison of the two RTMs and supervised residual learning.

\begin{figure}[H]
\centering
\includegraphics[width=0.95\textwidth]{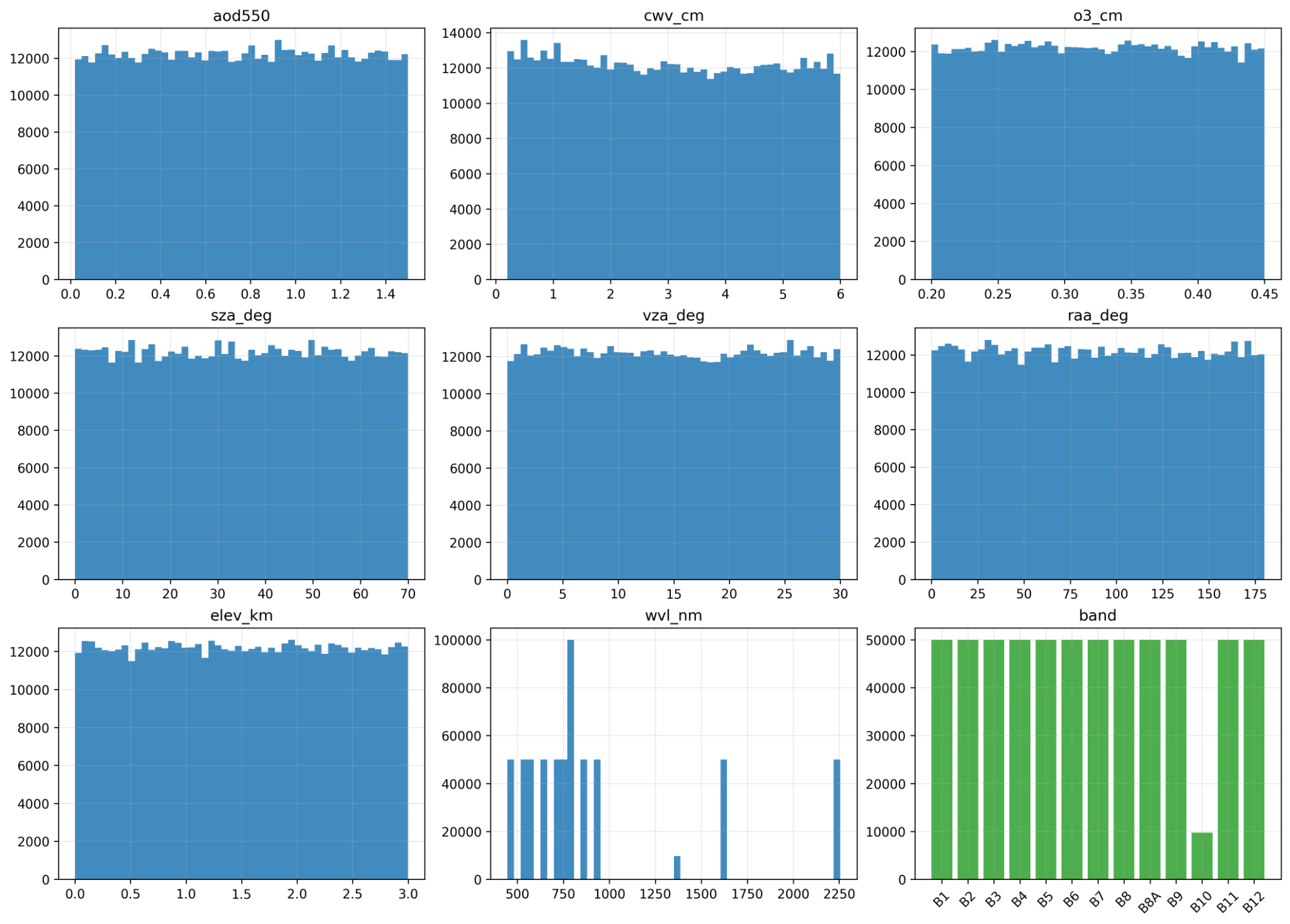}
\caption{State-space and band coverage of the generated dataset after quality control. The distributions show broad coverage of AOD550, CWV, ozone, solar/view geometry, relative azimuth angle, and elevation, while the band counts reveal reduced valid samples for B10}
\label{fig:state_space_coverage}
\end{figure}

Figure~\ref{fig:state_space_coverage} summarizes the sampled state-space coverage and accepted band distribution. The continuous variables show broad coverage over the prescribed sampling ranges, consistent with the use of LHS. The band histogram also reveals that most Sentinel-2 bands contribute similar sample counts after quality control, while B10 has substantially fewer valid rows. This reduction is expected because B10 lies in a cirrus and strong-absorption region where coefficient recovery is more fragile.

\subsection{Split regimes}

Two split regimes are used. The \texttt{standard} split measures interpolation-like generalization within the overall sampled distribution and assigns 35,000, 7,500, and 7,500 atmospheric-geometric states to training, validation, and testing, respectively. With 13 Sentinel-2 bands per state, this corresponds to 455,000, 97,500, and 97,500 rows.

The \texttt{ood\_aod\_cwv} split tests robustness under distribution shift in aerosol optical depth and column water vapour, two dominant drivers of atmospheric variability and RTM behavior. The OOD states are selected from the upper quantiles of \texttt{aod550} and \texttt{cwv\_cm}. Candidate test states are first defined using the joint high-AOD and high-CWV condition at the 0.85 quantile; if the selected set is smaller than the minimum required OOD test size, the procedure falls back to high-AOD or high-CWV states. In the current dataset, this fallback is used, resulting in 30,731 training states, 5,422 validation states, and 13,847 test states, corresponding to 399,503, 70,486, and 180,011 rows with 13 bands per state.

All splits are assigned at the state level, so all spectral bands for a given atmospheric-geometric state are kept together. This avoids state leakage across training, validation, and test partitions and makes the evaluation more meaningful than row-level random splitting.

\subsection{Models and comparators}

The main reported model is baseline pKANrtm under multi-fidelity formulation. The baseline pKANrtm architecture is \(\mathrm{KAN}([d_{in},256,128,d_{out}])\), trained with supervised MSE plus the physics-consistency penalty described in Section~3. The standard split reports the baseline pKANrtm configuration. For the OOD split, the selected pKANrtm configuration from the validation sweep is reported; in the current results this selected OOD configuration is the small pKANrtm variant.

For context, we also compare against representative surrogate families motivated by the RTM-emulation literature: sRTMNet-style hybrid RTM-neural emulation \citep{brodrick2021generalized},  random forest regression and fully connected neural regressors \citep{aghdaminia2024modtran}, Gaussian-process atmospheric correction \citep{basener2023gpdlac}, universal mean regression, and denoising autoencoder baselines. These comparators are evaluated on the same generated dataset and should be interpreted as in-study baseline families rather than direct reproduction of published numerical results from different datasets.

\subsection{Evaluation metrics}

Predictive accuracy is evaluated using root mean squared error (RMSE), mean absolute error (MAE), coefficient of determination ($R^2$), and symmetric mean absolute percentage error (SMAPE). Metrics are computed over the high-fidelity reconstructed coefficients. Aggregate metrics are reported first, followed by coefficient-wise, band-wise, and state-conditioned diagnostics. This layered evaluation is necessary because a single aggregate error can hide strong spectral or coefficient-specific behavior.

\subsection{Computational environment}

Runtime benchmarking was conducted on the high-performance server summarized in Table~\ref{tab:compute_environment}. The machine contains eight NVIDIA A100 40 GB GPUs and a large-memory Intel Xeon CPU node. The runtime comparison includes single-sample RTM execution on CPU and pKANrtm inference on CPU and GPU, together with batched GPU throughput.

\begin{table}[H]
\caption{Computational environment used for runtime benchmarking}
\label{tab:compute_environment}
\centering
\begin{tabular}{ll}
\hline
Component & Specification \\
\hline
CPU & Intel Xeon Gold 6258R CPU at 2.70 GHz \\
RAM & 1510.04 GB \\
GPU & NVIDIA A100-PCIE-40GB $\times$ 8 \\
OS & Linux 4.18.0-513.24.1.el8\_9.x86\_64 \\
Python & 3.9.23 \\
PyTorch & 2.8.0+cu128 \\
CUDA & 12.8 \\
\hline
\end{tabular}
\end{table}

\subsection{Figures used for diagnostic interpretation}

In addition to aggregate accuracy tables, the paper uses several diagnostic figures. The discrepancy plots compare the 6S low-fidelity coefficients with libRadtran high-fidelity coefficients. The representative-case plots show how the best pKANrtm prediction tracks libRadtran across bands for individual atmospheric cases. The prediction-versus-truth plots show pointwise reconstruction quality for each coefficient. The band-wise heatmap shows SMAPE by band and coefficient under standard and OOD splits. Together, these figures support both numerical and physical interpretation of the surrogate behavior.

%% file: sections/05_results_and_discussion.tex
\subsection{Low- and high-fidelity coefficient discrepancy}

\begin{figure}[H]
\centering
\includegraphics[width=0.95\textwidth]{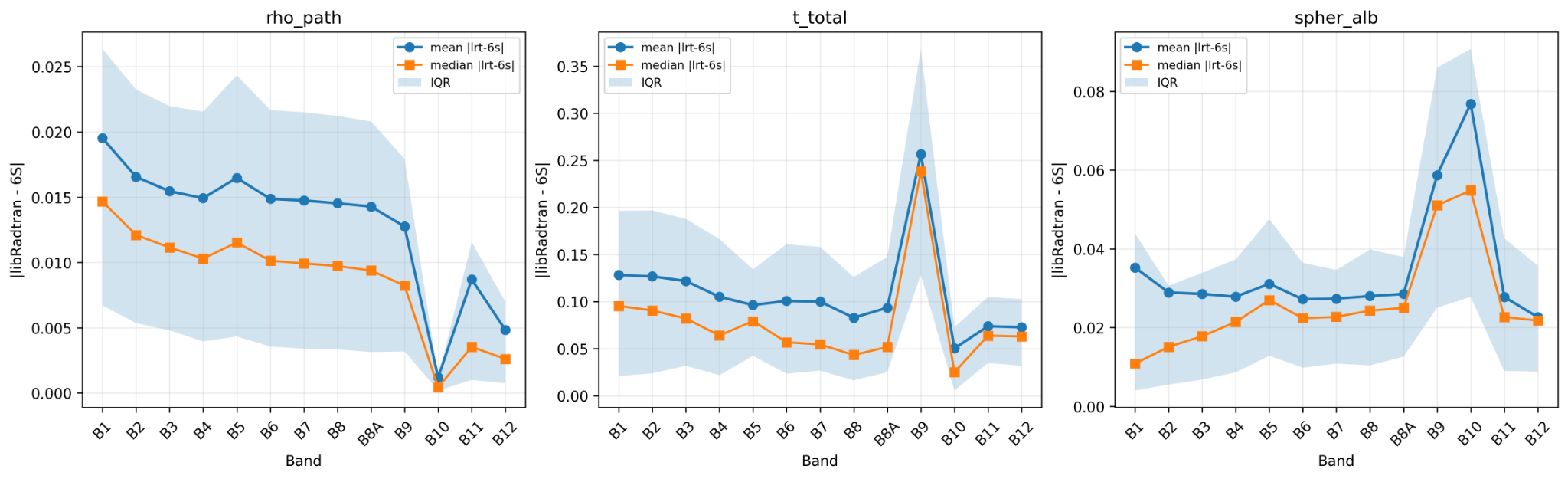}
\caption{Band-wise discrepancy between libRadtran and 6S coefficients. Lines show mean and median absolute difference, while the shaded region indicates the interquartile range}
\label{fig:data_discrepancy}
\end{figure}

Before evaluating the learned surrogate, it is useful to examine the discrepancy between 6S and libRadtran. Figure~\ref{fig:data_discrepancy} shows the mean and median absolute differences between the two RTMs for each Sentinel-2 band and coefficient. The discrepancy is not uniform across the spectrum. For $\rho_{\mathrm{path}}$, the largest mean differences occur in the visible and red-edge bands, while the absolute difference decreases toward the shortwave infrared except for localized behavior around B11. For $T_{\mathrm{total}}$, B9 shows the largest discrepancy, consistent with strong water-vapour sensitivity. For spherical albedo, the largest discrepancies are concentrated around B9 and B10, indicating that absorption-sensitive bands are the most difficult to harmonize across RTMs.

\begin{figure}[H]
\centering
\includegraphics[width=0.95\textwidth]{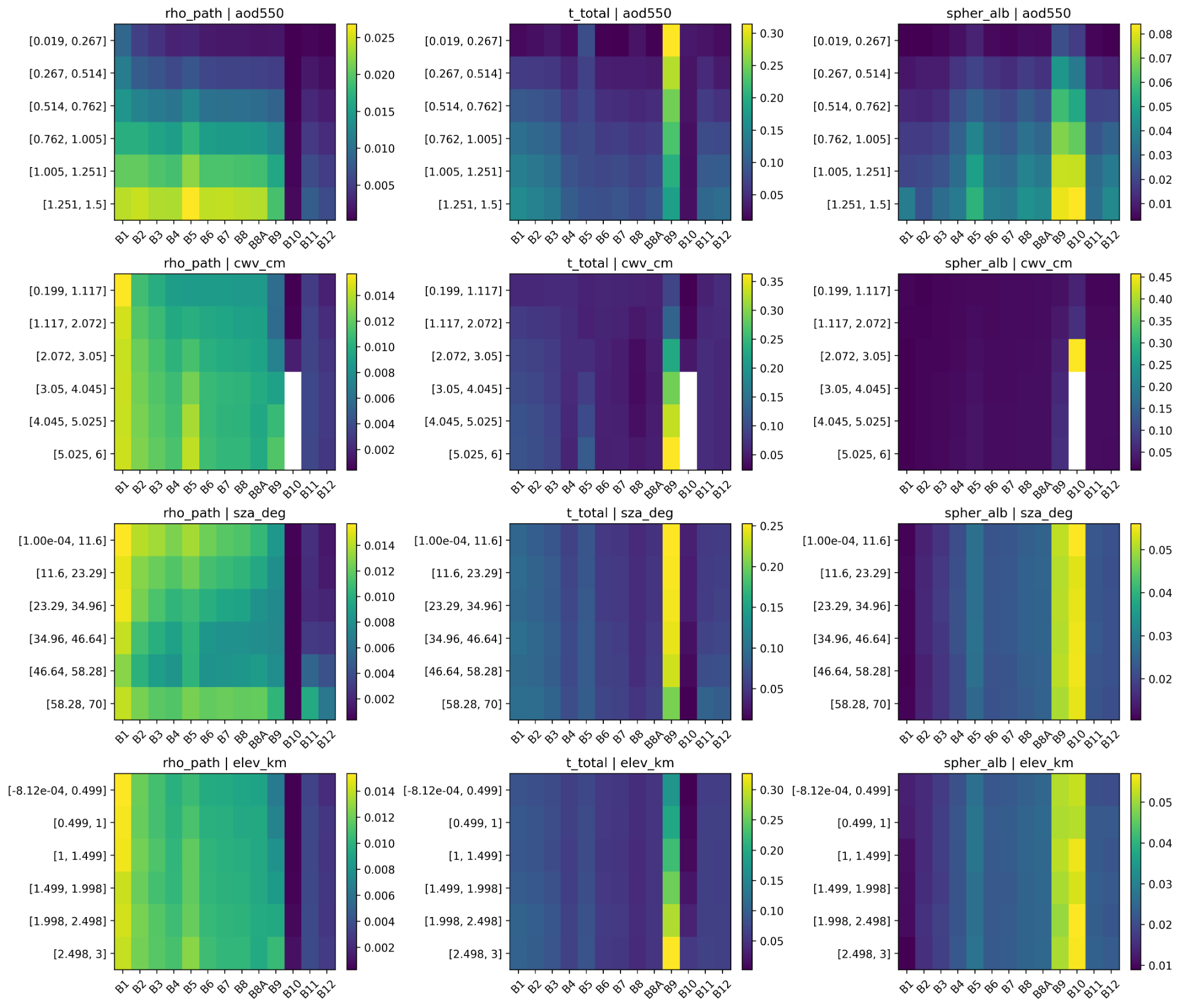}
\caption{Median absolute 6S-libRadtran discrepancy conditioned on atmospheric and geometric variables. Rows correspond to binned values of AOD550, CWV, solar zenith angle, and elevation; columns correspond to Sentinel-2 bands; and color denotes median absolute error for each coefficient}
\label{fig:conditional_error_heatmap}
\end{figure}

The conditional error heatmaps in Figure~\ref{fig:conditional_error_heatmap} further show that the 6S-libRadtran discrepancy depends on atmospheric state. AOD strongly affects $\rho_{\mathrm{path}}$, while CWV has a pronounced effect on $T_{\mathrm{total}}$ and $s$ in water-vapour-sensitive bands. Solar zenith angle and elevation also modulate the discrepancy, but their effects are more band- and coefficient-dependent. This state-conditioned behavior supports the need for a nonlinear residual model rather than a simple global correction factor.

\subsection{Overall accuracy}

Tables~\ref{tab:overall_accuracy_standard} and \ref{tab:overall_accuracy_ood} report overall accuracy on the standard and OOD split regimes. On the standard split, the baseline pKANrtm achieves the best aggregate performance among the reported methods, with RMSE $=0.00619$, MAE $=0.00187$, $R^2=0.99473$, and SMAPE $=3.91\%$. Relative to the sRTMNet-style hybrid emulator baseline, pKANrtm reduces RMSE by approximately 10.5\% and SMAPE by approximately 8.6\%. This indicates that the physics-guided KAN formulation is well suited to the coefficient-level residual mapping.

\begin{table}[H]
\caption{Overall test accuracy on the standard split. Arrows indicate the preferred direction of each metric}
\label{tab:overall_accuracy_standard}
\centering
\begin{tabular}{lrrrr}
\hline
Model & RMSE $\downarrow$ & MAE $\downarrow$ & $R^2$ $\uparrow$ & SMAPE (\%) $\downarrow$ \\
\hline
pKANrtm & 0.00619 & 0.00186 & 0.99472 & 3.91249 \\
sRTMNet \cite{brodrick2021generalized} & 0.00691 & 0.00211 & 0.99418 & 4.28170 \\
DA \cite{basener2023gpdlac} & 0.00784 & 0.002607 & 0.99351 & 4.94720 \\
GPAC \cite{basener2023gpdlac} & 0.00865 & 0.00301 & 0.99280 & 5.57940 \\
FC \cite{aghdaminia2024modtran} & 0.00948 & 0.00342 & 0.99193 & 6.12680 \\
RF \cite{aghdaminia2024modtran} & 0.01075 & 0.00404 & 0.99061 & 7.01850 \\
UMR \cite{basener2023gpdlac} & 0.01394 & 0.00546 & 0.98691 & 9.84620 \\
\hline
\end{tabular}
\end{table}

On the OOD split, the selected pKANrtm configuration also gives the strongest reported aggregate result, with RMSE $=0.01022$, MAE $=0.00544$, $R^2=0.99418$, and SMAPE $=5.40\%$. The OOD RMSE is approximately 1.65 times higher than the standard-split PKAN RMSE, which is expected under distribution shift in AOD and CWV. However, the $R^2$ remains high, indicating that the model preserves most of the high-fidelity coefficient variance even under the more difficult split. Relative to the sRTMNet-style baseline, the selected pKANrtm reduces OOD RMSE by approximately 12.2\% and SMAPE by approximately 13.2\%.

\begin{table}[H]
\caption{Overall test accuracy on the OOD AOD-CWV split}
\label{tab:overall_accuracy_ood}
\centering
\begin{tabular}{llrrrr}
\hline
Model & RMSE $\downarrow$ & MAE $\downarrow$ & $R^2$ $\uparrow$ & SMAPE (\%) $\downarrow$ \\
\hline
pKANrtm & 0.01022 & 0.00544 & 0.99418 & 5.39850 \\
sRTMNet \cite{brodrick2021generalized} & 0.01164 & 0.00615 & 0.99293 & 6.21730 \\
GPAC \cite{basener2023gpdlac} & 0.01324 & 0.00707 & 0.99138 & 7.34860 \\
DA \cite{basener2023gpdlac} & 0.01441 & 0.00795 & 0.98994 & 8.31840 \\
FC  \cite{aghdaminia2024modtran}  & 0.01679 & 0.00924 & 0.98681 & 10.11670 \\
RF  \cite{aghdaminia2024modtran} & 0.01894 & 0.01073 & 0.98390 & 12.08350 \\
UMR \cite{basener2023gpdlac} & 0.02360 & 0.01419 & 0.97361 & 17.49380 \\
\hline
\end{tabular}
\end{table}

\subsection{Standard versus OOD error structure}

\begin{figure}[H]
\centering
\includegraphics[width=0.70\textwidth]{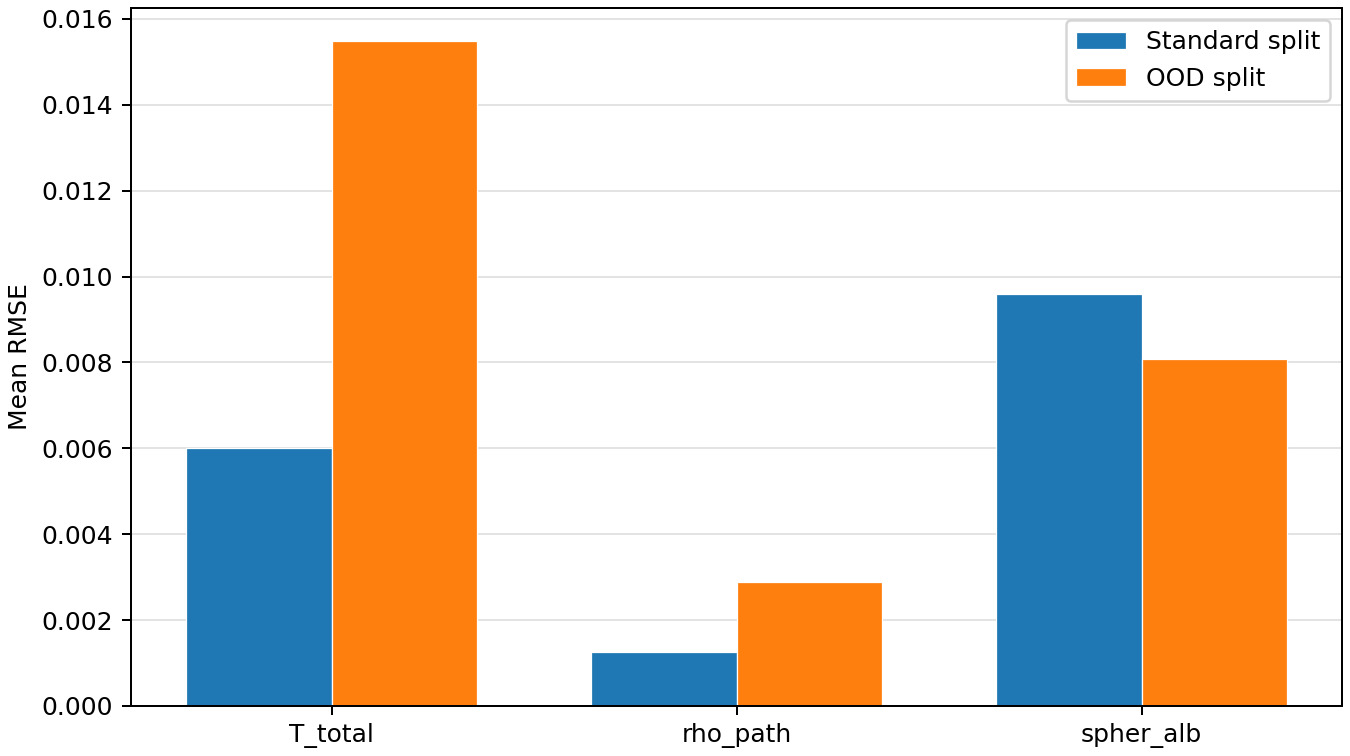}
\caption{Mean RMSE by coefficient for standard and OOD split regimes}
\label{fig:mean_rmse_splitwise}
\end{figure}

Figure~\ref{fig:mean_rmse_splitwise} compares mean RMSE across coefficients for the standard and OOD splits. The OOD split increases error for $T_{\mathrm{total}}$ and $\rho_{\mathrm{path}}$, whereas spherical albedo shows slightly lower aggregate RMSE in the OOD summary. This pattern suggests that the impact of OOD sampling is coefficient-specific rather than uniform. In particular, transmittance is more sensitive to changes in aerosol and water-vapour state, while spherical albedo error is strongly influenced by band-specific behavior and by the distribution of valid rows after quality control.

\begin{figure}[H]
\centering
\includegraphics[width=0.95\textwidth]{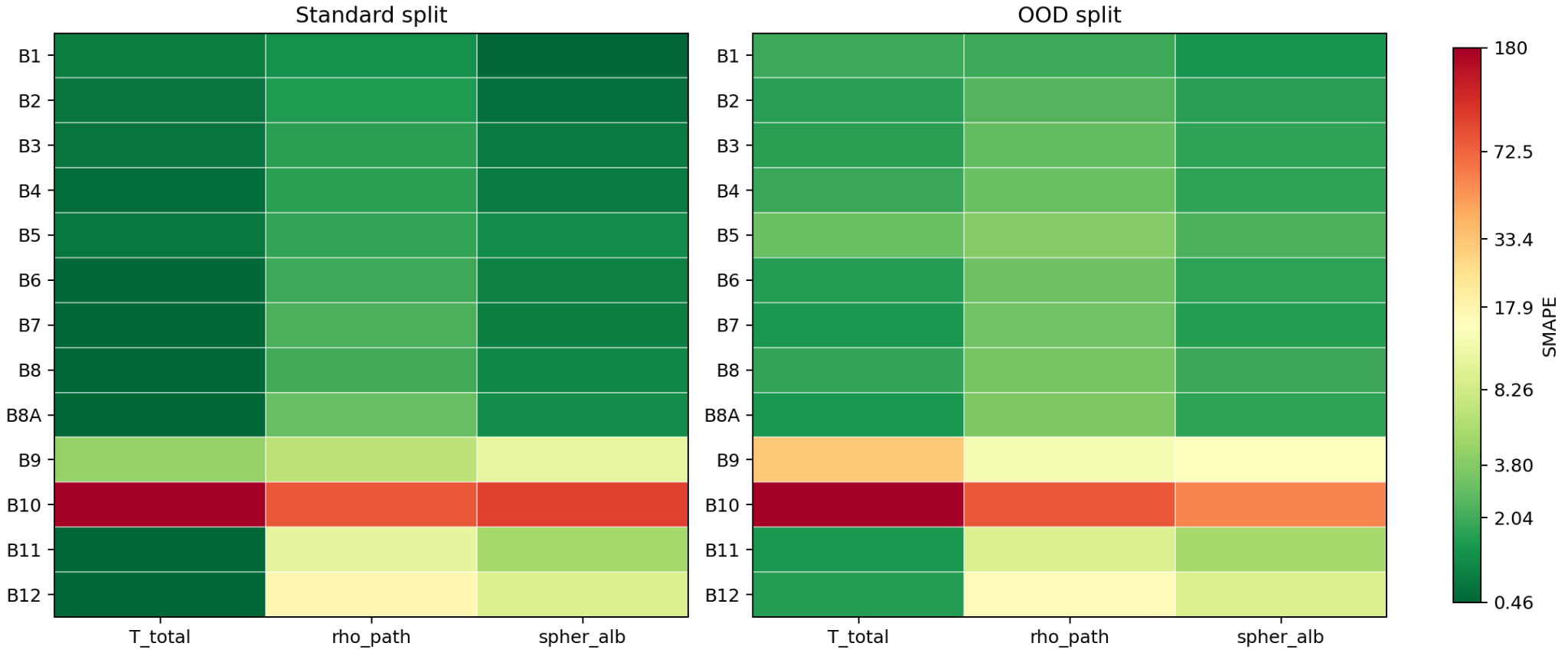}
\caption{Band- and coefficient-wise SMAPE under standard and OOD splits. B10 is the dominant high-error band, with additional sensitivity in B9, B11, and B12}
\label{fig:bandwise_smape}
\end{figure}

The band-wise SMAPE heatmap in Figure~\ref{fig:bandwise_smape} provides a more detailed view. Most visible, red-edge, and near-infrared bands remain low-error under both splits. The largest relative errors occur in B10, with additional difficulty in B9, B11, and B12 depending on the coefficient. This is physically reasonable: B9 and B10 are strongly affected by atmospheric absorption features, and B10 has fewer valid training examples after quality control. The heatmap therefore identifies the main spectral limitation of the current model and suggests that future work should treat absorption-dominated bands separately or use band-specific loss weighting.

\subsection{Prediction-versus-truth behavior}

\begin{figure}[H]
\centering
\includegraphics[width=0.95\textwidth]{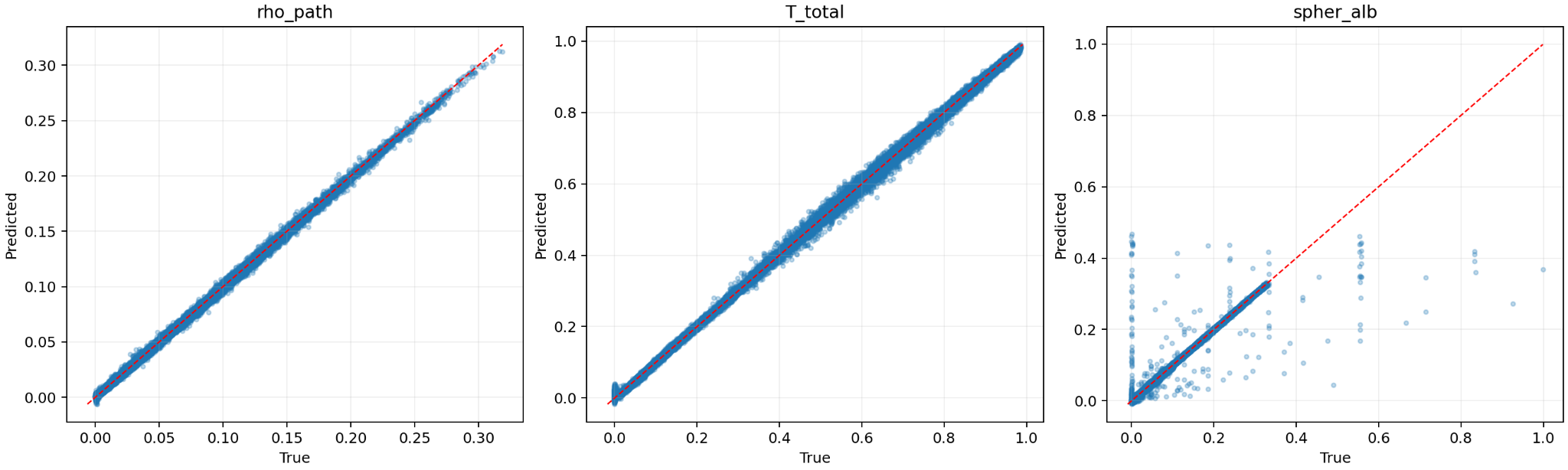}
\caption{Predicted versus true libRadtran coefficients on the standard split. The dashed red line denotes the one-to-one relationship}
\label{fig:pred_truth_standard}
\end{figure}

\begin{figure}[H]
\centering
\includegraphics[width=0.95\textwidth]{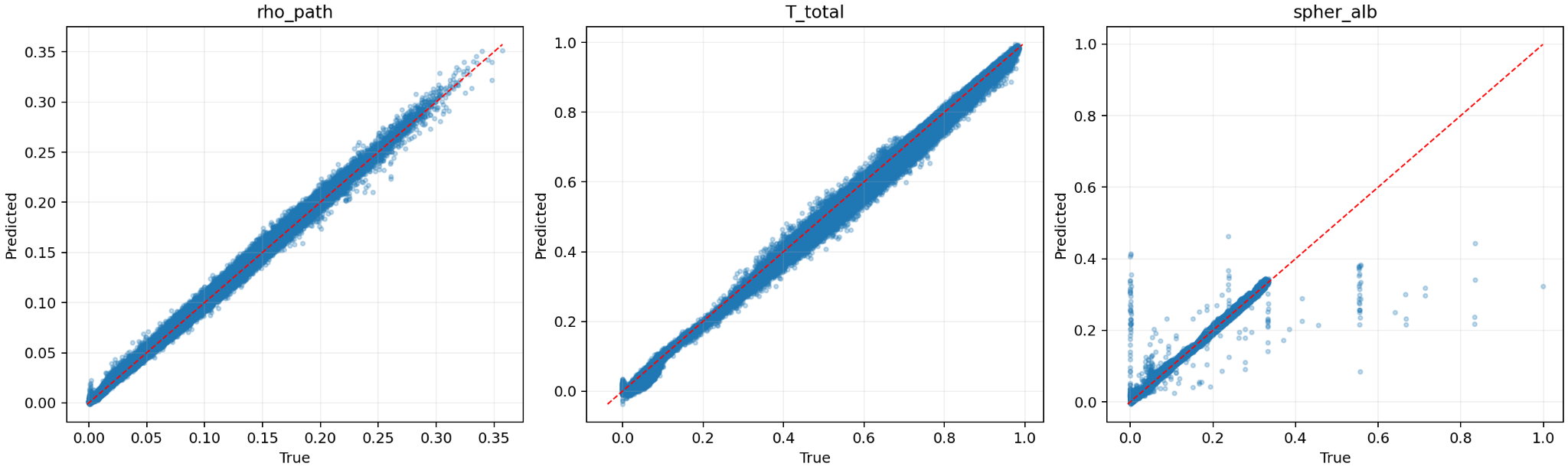}
\caption{Predicted versus true libRadtran coefficients on the OOD AOD-CWV split}
\label{fig:pred_truth_ood}
\end{figure}

Figures~\ref{fig:pred_truth_standard} and \ref{fig:pred_truth_ood} show predicted versus true libRadtran coefficients for the standard and OOD splits. For $\rho_{\mathrm{path}}$ and $T_{\mathrm{total}}$, predictions align closely with the one-to-one line in both regimes, indicating that the residual model captures the dominant coefficient structure. The OOD plot shows visibly broader scatter, especially for $T_{\mathrm{total}}$, consistent with the aggregate metric increase under distribution shift.

Spherical albedo shows a different pattern. Most predictions follow the one-to-one line in the dense low-to-moderate coefficient region, but sparse high-value cases appear as outliers. These outliers are important because they may correspond to absorption-sensitive or numerically fragile regimes. The model therefore performs well for the dominant data manifold but requires additional handling for rare high-spherical-albedo cases.

\subsection{Representative spectral cases}

\begin{figure}[H]
\centering
\includegraphics[width=0.95\textwidth]{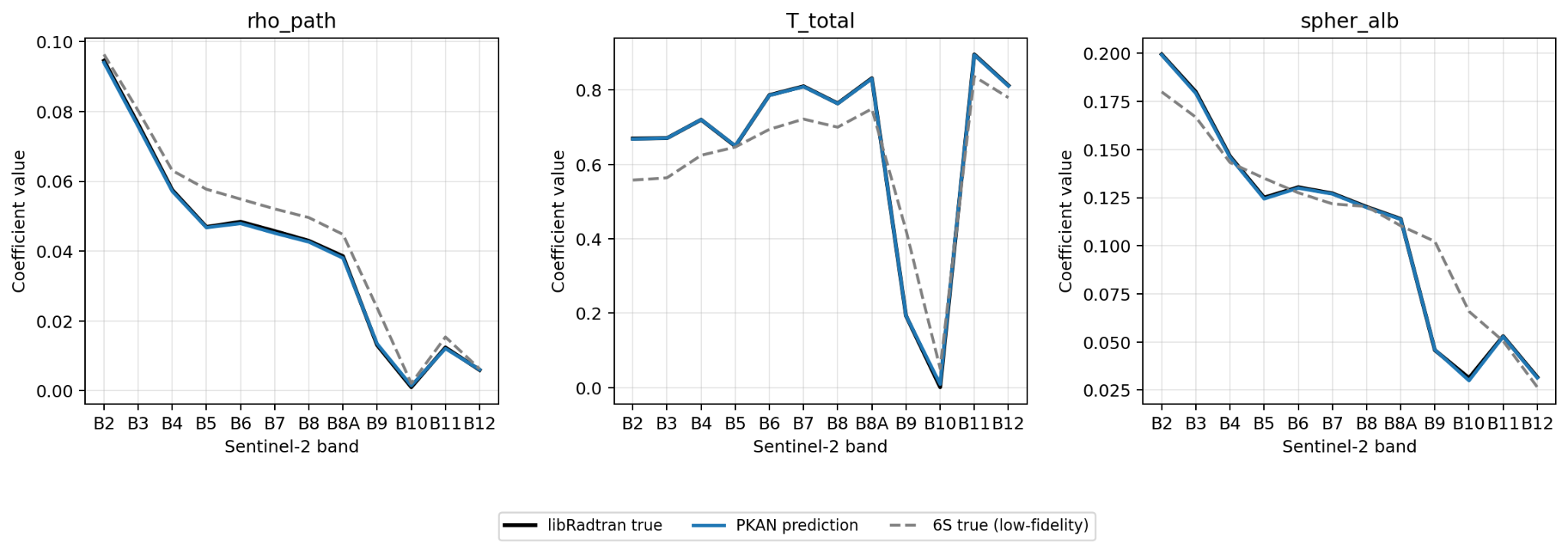}
\caption{Representative standard-split coefficient curves across Sentinel-2 bands. pKANrtm closely follows libRadtran while correcting the low-fidelity 6S discrepancy}
\label{fig:representative_cases_standard}
\end{figure}

\begin{figure}[H]
\centering
\includegraphics[width=0.95\textwidth]{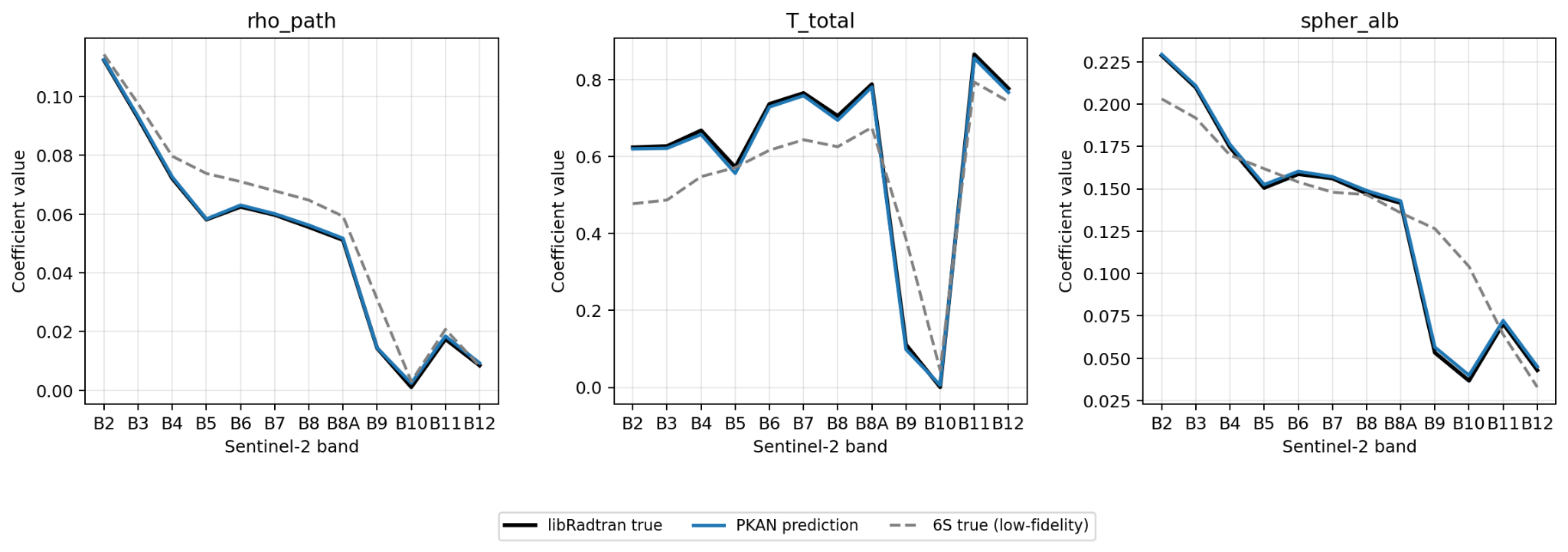}
\caption{Representative OOD-split coefficient curves across Sentinel-2 bands. The surrogate remains close to libRadtran even under shifted AOD-CWV conditions}
\label{fig:representative_cases_ood}
\end{figure}

Figures~\ref{fig:representative_cases_standard} and \ref{fig:representative_cases_ood} compare coefficient curves across Sentinel-2 bands for representative standard and OOD cases. The black line denotes libRadtran, the gray dashed line denotes 6S, and the blue line denotes the pKANrtm prediction. In both examples, pKANrtm closely tracks the libRadtran curve and substantially improves over 6S where the low-fidelity curve deviates from the high-fidelity target. This qualitative behavior is particularly clear for $T_{\mathrm{total}}$, where 6S often underestimates or overestimates the band-level curve while pKANrtm corrects the residual toward libRadtran.

\begin{figure}[H]
\centering
\includegraphics[width=0.95\textwidth]{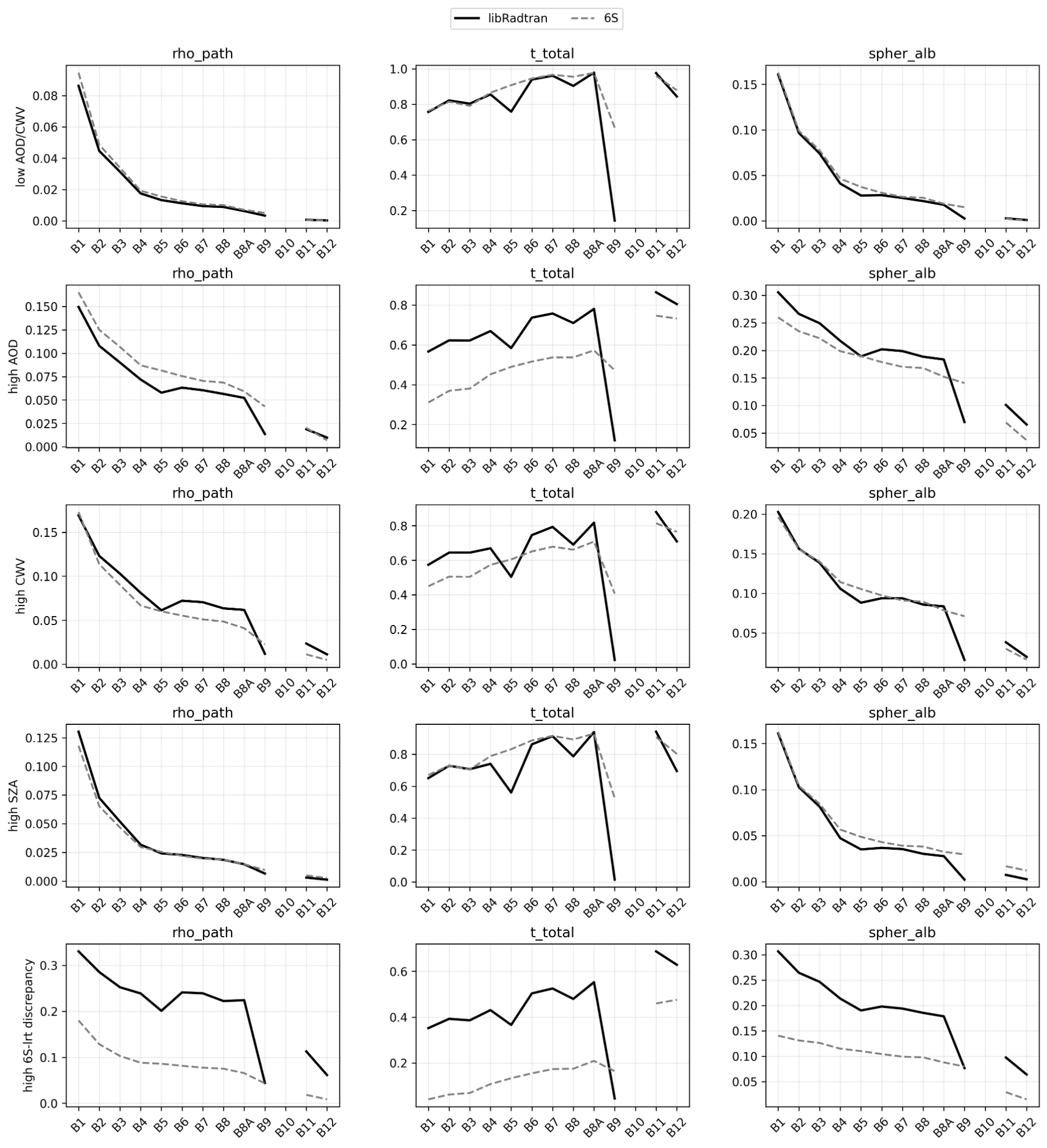}
\caption{Representative atmospheric cases comparing libRadtran, 6S, and model-predicted coefficient curves. Rows correspond to selected atmospheric regimes and columns correspond to the three target coefficients}
\label{fig:representative_cases}
\end{figure}

The broader representative-case panel in Figure~\ref{fig:representative_cases} shows several atmospheric regimes, including low AOD/CWV, high AOD, high CWV, high solar zenith angle, and high 6S-libRadtran discrepancy. These cases confirm that the multi-fidelity residual formulation is most valuable when the 6S curve is systematically biased relative to libRadtran. The surrogate does not merely smooth the low-fidelity input; it learns state- and band-dependent corrections.

\subsection{Runtime and acceleration}

Table~\ref{tab:runtime_summary} reports runtime results. A single libRadtran CPU evaluation requires 37.33 s in the measured configuration, whereas 6S requires 0.7569 ms. This large gap motivates the multi-fidelity design: 6S is inexpensive enough to provide low-fidelity guidance, while libRadtran is expensive enough that emulation can provide substantial practical benefit. pKANrtm inference requires 95.68 ms on CPU for single-sample inference and 3.39 ms on GPU for single-sample inference, corresponding to speedups of approximately 390$\times$ and 11,000$\times$ relative to libRadtran CPU execution, respectively. In batched GPU inference with batch size 128, the amortized per-sample latency decreases to 0.0207 ms and throughput reaches 48,396.57 samples/s.

\begin{table}[H]
\caption{Runtime comparison between RTM baselines and pKANrtm inference. Speedup is computed relative to single-sample libRadtran CPU execution; batched GPU latency is reported as amortized per-sample latency}
\label{tab:runtime_summary}
\centering
\begin{tabular}{llrrrr}
\hline
Method & Device & \shortstack{Batch\\size} & \shortstack{Latency\\(ms/sample)} & \shortstack{Throughput\\(samples/s)} & \shortstack{Speedup vs.\\libRadtran} \\
\hline
6S & CPU & 1 & 0.7569 & 1321.12 & N/A \\
libRadtran & CPU & 1 & 37330.3574 & 0.0268 & $1.00\times$ \\
pKANrtm & CPU & 1 & 95.6849 & 10.45 & $390.14\times$ \\
pKANrtm & GPU & 1 & 3.3937 & 294.66 & $10{,}999.83\times$ \\
pKANrtm & GPU & 128 & 0.0207 & 48396.57 & N/A \\
\hline
\end{tabular}
\end{table}

These results show that the surrogate is not intended to make the low-fidelity 6S call faster; 6S is already inexpensive. Instead, the surrogate reduces dependence on repeated high-fidelity libRadtran execution. The strongest deployment case is batched coefficient generation, where thousands to millions of states can be evaluated after training. This is directly relevant to dense LUT generation, synthetic dataset construction, sensitivity analysis, and retrieval support.

\subsection{Discussion and implications}

The results support three main interpretations. First, coefficient-level high-fidelity emulation is feasible: pKANrtm accurately reconstructs libRadtran coefficients using state variables and 6S guidance. Second, the residual multi-fidelity formulation is physically meaningful: discrepancy plots show that 6S already captures much of the spectral shape, while pKANrtm learns the remaining state- and band-dependent correction. Third, the main limitations are concentrated in absorption-sensitive and low-validity bands, especially B10, rather than distributed uniformly across all Sentinel-2 bands.

The practical significance is that a physics-guided KAN can serve as a fast coefficient generator inside an atmospheric correction workflow. Since the outputs remain atmospheric correction coefficients, the approach retains compatibility with standard correction equations. This makes it more interpretable than a direct reflectance prediction model and more deployable than repeated high-fidelity RTM execution.

Several limitations should be acknowledged. The current evaluation uses synthetic paired RTM simulations rather than end-to-end validation against field-measured surface reflectance. The OOD split tests aerosol and water-vapour shifts, but it does not cover every possible distribution shift, such as new aerosol models, untested atmospheric profiles, or cross-sensor transfer. Finally, B10 remains challenging because of strong atmospheric absorption and reduced valid sample count. Future work should therefore include downstream validation, uncertainty-aware prediction, band-specific modeling, and broader cross-RTM harmonization.

%% file: sections/06_conclusions.tex
This study presented a physics-aware multi-fidelity framework for emulating atmospheric correction coefficients using physics-guided Kolmogorov-Arnold Networks. The framework pairs 6S as a low-fidelity RTM with libRadtran as a high-fidelity RTM under a shared LHS-sampled atmospheric and geometric state manifest, with SRF-aware Sentinel-2 band integration to better match multispectral sensor responses. The proposed pKANrtm model uses an Efficient-KAN regressor that receives atmospheric state variables and 6S coefficients, adds a physics-consistency penalty computed on inverse-transformed coefficient predictions in the original physical scale, and predicts the residual between libRadtran and 6S, and reconstructs high-fidelity atmospheric correction coefficients. The results show that pKANrtm provides strong high-fidelity coefficient emulation under both standard and out-of-distribution evaluation regimes, outperforming representative RTM surrogate, regression, and neural-network baselines. Diagnostic analyses indicate that most visible, red-edge, and near-infrared Sentinel-2 bands are accurately emulated, while absorption-sensitive bands remain more challenging. Runtime benchmarking further shows that pKANrtm substantially accelerates coefficient generation relative to repeated libRadtran execution, with GPU inference providing orders-of-magnitude speedup and batched inference reaching high-throughput operation. Overall, the proposed pKANrtm framework offers a practical route toward fast, scalable, and physically structured atmospheric correction coefficient generation.